\documentclass[aps,prb,amsmath,amssymb,superscriptaddress,floatfix,showpacs,twocolumn]{revtex4}

\usepackage{graphicx} 
\usepackage{bm}
\usepackage{hyperref}

\begin{document}

\title{Supersolid phase and magnetization plateaus observed in anisotropic spin--$3/2$ Heisenberg model on bipartite lattices}

\author{Judit Romh\'anyi}
\affiliation{Research Institute  for  Solid State  Physics and Optics, H--1525 Budapest, P.O.B.~49, Hungary}
\affiliation{Department of Physics, Budapest University of Technology and Economics, H--1111 Budapest, Budafoki \'ut 8, Hungary}

\author{Frank Pollmann}
\affiliation{Max-Planck-Institut f{\"u}r Physik komplexer Systeme, 01187 Dresden, Germany}

\author{Karlo Penc}
\affiliation{Research Institute  for  Solid State  Physics and Optics, H--1525 Budapest, P.O.B.~49, Hungary}

\date{\today}

\begin{abstract}   
We study the spin-$3/2$ Heisenberg model including easy--plane and exchange anisotropies in one and two dimensions. 
In the Ising limit, when the off--diagonal exchange interaction $J$ is zero, the phase diagram in magnetic field is characterized by magnetization plateaus that are either translationally invariant or have a two--sublattice order, with phase boundaries that are macroscopically degenerate.
Using a site factorized variational wave function and perturbational expansion around the Ising limit, we find that superfluid and supersolid phases emerge between the plateaus for small finite values of $J$.
 The variational approach is complemented by a Density Matrix Renormalization Group study of a one-dimensional chain and exact diagonalization calculations on small clusters of a square lattice. The studied model may serve as a minimal model for the layered Ba$_2$CoGe$_2$O$_7$ material compound, and we believe that the vicinity of the uniform 1/3 plateau in the model parameter space can be observed as an anomaly in the measured magnetization curve. 
\end{abstract}
\maketitle

\section{Introduction and the model}
Finding systems -- both theoretically and experimentally -- that exhibit novel quantum phases, amongst them supersolid states, played an important role in the study of strongly correlated systems in the last fifty years. Superfluid (as well as superconducting) phases and quantum crystals can be characterized by off-diagonal long range order (ODLRO)\cite{Yang1962} and diagonal long range oder (DLRO) respectively. This classification allows us to think about supersolid phases as states in which ODLRO and DLRO coexists. 
Supersolid phases were first observed in the context of strongly interacting bosons of ${^4}$He that can simultaneously Bose condensed and order in crystalline solid.\cite{Chester1970,Leggett1970,Andreev1971} Experimental evidence\cite{Kim2004,Rittner2006,Kondo2007,Aoki2007} of the existence of such phase was found after almost half a century, reviving the theoretical interest in supersolid states, and indicating that theoretical interpretation might be more difficult than the first ideas.\cite{Leggett2004,Anderson2007,Anderson2008,Anderson2009} 

Apparently various bosonic lattice models provide a better understanding of supersolid phases. 
Quantum Monte Carlo (QMC) simulations for hard-core bosonic Hubbard model on square lattice with nearest-neighbor and next-nearest-neighbor interactions suggested that the 'checkerboard' supersolid phase is thermodynamically unstable, however -- through continuous phase transition from  the superfluid state -- a stable 'striped' supersolid emerges.\cite{Batrouni2000} Similar QMC simulations of a soft-core boson model of square lattice indicated a supersolid phase that is stable against phase separation.\cite{Sengupta2005,PhysRevB.79.094503} 

Matsuda and Tsuneto, and Liu and Fisher showed that the bosonic picture of supersolid state can be mapped onto a model of magnetic supersolid where the magnetic order breaks spin rotational symmetry and translational invariance at the same time.\cite{Matsuda1970,Liu1973} Such magnetic analogs of supersolid state were observed in triangular lattice via Quantum Monte Carlo (QMC) simulations\cite{Wessel2005,Melko2005,PhysRevLett.95.127206} where frustration and order-by-disorder mechanism plays an important role in the emergence of supersolid phase. Classical Monte Carlo simulation on triangular lattice supported by mean-field calculation and Landau theory suggested that strong anisotropy can stabilize supersolid phases.\cite{Seabra2011} Amongst quasi two dimensional systems QMC simulations on bilayer dimer models\cite{Ng2006,Sengupta2007,Laflorencie2007,Picon2008} and orthogonal dimer models\cite{Schmidt2008} were also found to exhibit supersolid states that are stabilized by strong frustration and/or anisotropy. 
Supersolid states have also been reported in the spin-$1$ Heisenberg chain with strong exchange and uniaxial single--ion anisotropies.\cite{Sengupta2007/2,Peters2009,Peters2010,Rossini2011} Furthermore a supersolid phase was found in three dimensional spin and hard-core Bose-Hubbard model as well.\cite{Ueda2010}

In this paper we investigate spin-$3/2$  (quantum) antiferromagnetic models on a square lattice and on a chain with both easy--plane and exchange anisotropies described by the following Hamiltonian:
\begin{eqnarray}
\mathcal{H}&=&J\sum_{\langle i,j\rangle}\left(\hat S^x_i \hat S^x_j+\hat S^y_i \hat S^y_j\right)+J_z \sum_{\langle i,j\rangle}\hat S^z_i \hat S^z_j\nonumber\\
&&+\Lambda\sum_i \left(\hat S^z_i\right)^2 + h \sum_i \hat S^z_i 
\label{eq:Hamiltonian}
\end{eqnarray}
where $\langle i,j \rangle$ indicates nearest neighbor sites.
Our model is inspired by the quasi two dimensional  Ba$_2$CoGe$_2$O$_7$, where the magnetic spin--$3/2$ Co$^{2+}$ ions form layers of strongly anisotropic square lattices.\cite{Zheludev2003,Yi2008,Murakawa2010,Miyahara2011}


The paper is structured as follows:  In Sec.~\ref{sec:ising} we discuss the phase diagram in the Ising limit and the instabilities of the plateaus using perturbation theory. In the following section (Sec. \ref{sec:variational}) we map out the phase diagram using a variational approximation in different cases, and determine the stability of the plateaus and of the supersolid phases. To check the reliability of the variational method, we calculate the phase diagram for the spin--1 model and compare it to the known results in the literature. In Sec. \ref{sec:chain} a one dimensional chain is studied using a variant of the Density Matrix Renormalization Group method and evidence for the existence of an intermediate supersolid phase is presented. In Sec. \ref{sec:exactdiag} we show results of an exact diagonalization study on a square lattice. We conclude with Sec.~\ref{sec:concusions}.

\section{The Ising limit and  perturbational expansions around it}
\label{sec:ising}

\subsection{The Ising limit}
\begin{figure}[h]
\begin{center}
\includegraphics[width=7cm]{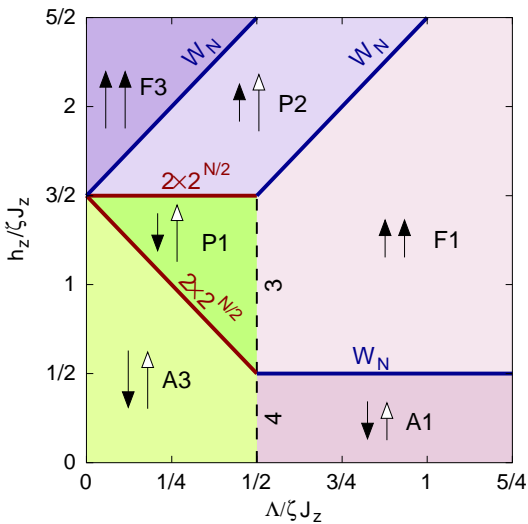}
\caption{(color online) Phase diagram in the Ising limit as the function of the anisotropy and magnetic field. The spin configurations on  A and B sublattice are shown, as well as the degeneracies of the ground state manifolds on the phase boundaries (the dashed line is a first order phase boundary). Long arrows represent the $S^z=\pm 3/2$ spin states, while the sort ones the $S^z = \pm 1/2$'s. The coordination number $\zeta=2$ for the chain and $\zeta=4$ for the square. $F1$ and $F3$ are uniform phases, while the others break the translational invariance and are two--fold degenerate.}
\label{fig:ising}
\end{center}
\end{figure}

The existence of the gapped phase in our model is due to the anisotropic terms, so turning off the $\hat S^x_i \hat S^x_j+\hat S^y_i \hat S^y_j$ off-diagonal Heisenberg term what remains are the plateaus. For brevity, we call this $J \rightarrow 0$ limit the Ising limit. Since the lattice is bipartite and we have nearest neighbor interactions only, the spins are not frustrated and all the ground states in the plateaus are either uniform or two-sublattice ordered. The ground state wave functions and their properties are listed in Table~\ref{tab:ising_wavef_orderp}, and the phase diagram as a function of magnetic field and single-ion anisotropy is outlined in Figure \ref{fig:ising}.

 Two uniform phases appear in finite magnetic field: the fully saturated state with $S^z=+3/2$ on each site and the $m/m_{\text{sat}}=1/3$ plateau state with $S_z=+1/2$ on the sites. We denote these states as $F3$ and $F1$, respectively.  The two--sublattice states include the two antiferromagnetic Ising--like states $A3$ and $A1$ with staggered magnetization $|S^z_A-S^z_B|=3$ and $|S^z_A-S^z_B|=1$ and vanishing uniform magnetization. In addition we find two other plateaus, $P1$ and $P2$, with magnetization that is $1/3$ and $2/3$ of the saturation magnetization, respectively. 
 
  The phase boundaries between different phases are established by comparing the ground state energies. A first order phase transition occurs between the $A1$ and $A3$ phases at $\Lambda = \zeta J_z/2$, when the lowest lying energy levels cross. ($\zeta$ stands for the coordination number.) The ground state degeneracy (4) at the phase boundary is just the sum of the degeneracy of the phases it separates (2+2). Since the other states are separated by a gap, we expect that the level crossing will persist even for finite values of  $J$. The phase transition between the phases $P1$ and $F1$  is of similar kind. 
  
  The phase boundaries between two--sublattice $A3$ and $P1$ states is more interesting: the ground state at the phase boundary is macroscopically degenerate, and goes as $2\times 2^{N/2}$. This degeneracy is understood in the following way: as we cross the boundary by increasing the field, the $S^z=+3/2$ spins on the $B$ sublattice do not change, while the $S^z=-3/2$ spins become $S^z=-1/2$ on the $A$ sublattice. At the boundary, the energy of having an $-3/2$ or $-1/2$ is equal, thus they create the $2^{N/2}$ fold degenerate manifold ($N/2$ is the number of sublattice sites). The additional factor of 2 comes from the choice of the sublattice ($A$ or $B$). Turning on $J$, this degeneracy will immediately be lifted (we may think of a pseudospin--1/2 Heisenberg like effective model to describe this problem), and a gapless phase appears. The same scenario holds for the phase boundary between the phases $P1$ and $P2$. These phase boundaries are shown by thick red line in Fig.~\ref{fig:ising}.
  
  Lastly, we examine the phase boundary between the uniform and two--sublattice states. These phase boundaries are shown by thick blue lines in Fig.~\ref{fig:ising} and have a ground state degeneracy $W_N$. Let us concentrate on the boundary that separates $P2$ and $F3$. The allowed nearest neighbor configurations are $(+3/2,+3/2)$, $(+3/2,+1/2)$ and $(+1/2,+3/2)$, while the $(+1/2,+1/2)$ is not allowed. In the one dimensional chain this rule gives a degeneracy $W_N=F_{N-1}+F_{N+1}$, where $F_N$ is the $N$-th Fibonacci number ($W_2=3$, $W_4=7$, $W_6 = 18$, $W_8 = 47$, and so on). \cite{PhysRevLett.98.160409} In the case of square lattice, we cannot give an explicit formula for $W_N$, numerically we find $W_8=31$ for the 8--site cluster with $D_4$ symmetry and $W_{10}=68$ for the 10--site cluster with $C_4$ symmetry (the degeneracy depends on the shape of the cluster). 
     
  Starting from this phase diagram, we study the effect of the off--diagonal exchange $J$ below, using perturbation theory.

\begin{table}
\caption{(color online) Summary of ground states in the Ising limit. The relevant order parameters in the Ising limit are the magnetization $m_z=\frac{1}{2}(S^z_A+S^z_B)$ and the staggered magnetization $m^{\text{st}}_z=\frac{1}{2}|S^z_A-S^z_B|$. We denote the fully and partially polarized antiferromagnetic states by $A3$ and $A1$, the fully and partially polarized ferromagnetic phases by $F3$ and $F1$, and finally the plateau states by $P2$ and $P1$ corresponding to the $2/3$ and $1/3$ plateaus respectively. Although, the partially polarized ferromagnetic state $F1$ is a plateau with $m/m_{\text{sat}}=1/3$, we (prefer to) call it ferromagnetic state and refer to the plateaus as states that exhibit both finite $m_z$ and $m^{\text{st}}_z$. $\zeta$ is the coordination number of the (bipartite) lattice. %
\label{tab:ising_wavef_orderp}
}
\begin{center}
\begin{ruledtabular}
\begin{tabular}{lccccc}
\raisebox{0.5ex}[0pt]{$|S_A^z S_B^z\rangle$ } &
\raisebox{0.5ex}[0pt]{$E_{0}/N$} &
\raisebox{0.5ex}[0pt]{$m_z$} & 
\raisebox{0.5ex}[0pt]{$m^{\text{st}}_z$} &
\raisebox{0.5ex}[0pt]{$m_z/m_{\text{sat}}$} &
\raisebox{0.5ex}[0pt]{notation}\\ 
\hline
$|\downarrow\uparrow\rangle$ & $\frac{1}{4}\Lambda - \frac{1}{8} \zeta J_z$ & 0 & $1/2$ & $0$ & {A1}  \\ 
$|\Downarrow\Uparrow\rangle$ & $\frac{9}{4}\Lambda - \frac{9}{8} \zeta J_z$ & $0$ & $3/2$ & $0$ & {A3}  \\ 
$|\uparrow\uparrow\rangle$ & $\frac{1}{4}\Lambda + \frac{1}{8} \zeta J_z-\frac{1}{2}h$ & $1/2$ & $0$ & $1/3$ & {F1}  \\ 
$|\downarrow\Uparrow\rangle$ & $\frac{5}{4}\Lambda - \frac{3}{8} \zeta J_z-\frac{1}{2}h$ & $1/2$ & $1$ & $1/3$ & {P1}  \\ 
$|\uparrow\Uparrow\rangle$ & $\frac{5}{4}\Lambda + \frac{3}{8} \zeta J_z-h$ & $1$ & $1/2$ & $2/3$ & {P2}  \\
$|\Uparrow\Uparrow\rangle$ & $\frac{9}{4}\Lambda + \frac{9}{8} \zeta J_z-\frac{3}{2} h$ & $3/2$ & $0$ & $1$ & {F3}  \\  
\end{tabular}
\end{ruledtabular}
\end{center}
\end{table}

\subsection{Mapping to an effective $XXZ$ model}
\label{sec:mappingXXZ}

Sufficiently far from the $\Lambda=0$ and $h=0$ points, where we are essentially dealing with two types of spins only ($|\Uparrow\rangle$ and $|\uparrow\rangle$),  the F3--P2--F1 phase transitions can be mapped to an effective spin--1/2 model XXZ model:
\begin{equation}
 \mathcal{H}_{\text{eff}} =  \tilde J \sum_{i,j} \left( \sigma_i^x \sigma_j^x + \sigma_i^y \sigma_j^y + \tilde \Delta \sigma_i^z \sigma_j^z \right) - \tilde h \sum_{i} \sigma_i^z 
 \label{eq:Hameff}
\end{equation}
where the $\sigma^\alpha_i$ are the spin--1/2 operators on site $i$ that act on the Hilbert space made of the $|\tilde\uparrow\rangle$ and $|\tilde\downarrow\rangle$ effective spins. Selecting the mapping
$|\Uparrow\rangle,|\uparrow\rangle \rightarrow  |\tilde\uparrow\rangle,|\tilde\downarrow\rangle$ and comparing the matrix elements between the S=3/2 Hamiltonian (\ref{eq:Hamiltonian})
and the effective Hamiltonian (\ref{eq:Hameff}), we obtain the following parameters for the mapping:  
\begin{eqnarray}
\tilde \Delta &=& \frac{J_z}{3J}, \\
\tilde J &=& 3 J, \\
\tilde h &=& h-2\Lambda - \zeta J_z.
\end{eqnarray}
The Mapping is valid in leading order of the off--diagonal exchange.
In this case, the $P2$ phase corresponds to the Ising phase of the effective model, and the $F3$ and $F1$ phases to the saturated phases of effective Hamiltonian. Analogously, the mapping
$|\uparrow\rangle,|\downarrow\rangle \rightarrow  |\tilde\uparrow\rangle,|\tilde\downarrow\rangle$ 
leads to 
\begin{eqnarray}
\tilde \Delta &=& \frac{J_z}{4J}, \\
\tilde J &=&  4 J, \\
\tilde h &=& h ,
\end{eqnarray}
effective interaction terms, and the phases $A1$ and $F1$ correspond to the Ising and the saturated phases of the effective model, respectively. 

The effective XXZ model is in a gapped Ising phase for $\tilde \Delta >1$. Thus it becomes clear from our mapping that the phase $P2$ disappears once $J \agt J_z/3$ and the phase $A1$ when $J \agt J_z/4$, with the phase $F1$ surviving.

The XXZ--model has been extensively studied in the literature, and numerical methods find no trace of supersolids on  bipartite lattices. Instead, the zero magnetization gapped phase of the XXZ model ($P2$ in the mapping) is separated by a first order transition from the gapless superfluid phase.\cite{Masanori1997,Seiji2002} The phase separation can be prevented, e.g., by longer range diagonal exchanges\cite{Batrouni2000}. Likewise, the supersolid phase can also be stabilized by introducing second neighbor correlated hoppings (in the language of the equivalent hard--core boson problem), where the hopping on the second neighbor depends on the occupancy of the site along the hopping path.\cite{Sengupta2005,Picon2008} Such terms may arise in higher orders of perturbations theory, but even then the existence of the supersolids is a question of very delicate balance between different terms. 

The physics of the  transitions between $P2$ and $P1$, and $P1$ and $A3$ cannot be mapped to an XXZ model in simple terms. In that case we shall distinguish sites that can be occupied with spins in three different states. Since one of the states ($\Uparrow$) occupies one of the sublattices, and the two other states share the the other sublattice, the mechanism (see, e.g., Ref.~[\onlinecite{Sengupta2007}]) that leads to phase separation is suppressed and the formation of the supersolid is much more natural.

\subsection{Estimating the first order phase transitions}
From the Ising phase diagram we learned that the boundary between  $A1$ and $A3$ is of first order, corresponding to level-crossing in the energy spectrum that is otherwise gapped. We may assume that for not too big values of $J$ this holds as well, so that we can estimate the corrections to the phase boundary by comparing the ground state energies that is expanded in powers of $J$. The lowest order corrections appear in the second order:
\begin{eqnarray}
  \frac{E_{A1}}{N} &=& \frac{\Lambda}{4}  - \frac{\zeta J_z}{8}  -  \frac{2 \zeta J^2}{(\zeta-1) J_z} - \frac{9 \zeta J^2}{32 \Lambda - 8 (\zeta+1) J_z},  \label{eq:e2A1M} \\ 
  \frac{E_{A3}}{N} &=& \frac{9\Lambda}{4}  - \frac{9\zeta J_z}{8}  - \frac{9\zeta J^2}{(24\zeta -8) J_z - 32 \Lambda}. \label{eq:e2A3M}
 \end{eqnarray}
%
Comparing these energies, we get that the first order phase transition  between $A1$ and $A3$ in the square lattice happens when 
\begin{equation}
 \Lambda=2 J_z -\frac{4 J^2}{3 J_z} + O(J^4)
\end{equation} 
for small $J$. In the case of the one--dimensional chain we get 
\begin{equation}
 \Lambda= J_z -\frac{2 J^2}{J_z} + O(J^4).
\end{equation} 

Similarly, from the second order corrections given in the Appendix, Eqs.~(\ref{eq:e2P1}) and (\ref{eq:e2F1}), the boundary between the phases P1 and F1 is 
\begin{eqnarray}
 \Lambda &=& 2 J_z -\frac{2 J^2}{J_z} + O(J^4) ,
 \end{eqnarray} 
 for a square lattice and
 \begin{eqnarray}
 \Lambda &=& J_z -\frac{3 J^2}{J_z} + O(J^4) ,
\end{eqnarray} 
for a chain.

\subsection{Field induced instability of uniform phases}

The field induced instability of Ising phases can be thought of as a softening of magnetic excitations. The simplest magnetic excitations corresponds to lowering or raising the spins on a site that becomes delocalized due to the off--diagonal $J$ term. These excitations are gapped in the Ising (plateau) phases, and the value of the gap changes with magnetic field and interaction parameters. When the energy gap vanishes, it means that these excitations can be created in arbitrary number and an off--diagonal long--range order develops. For small values of $J$ we can use perturbation expansion to get the dispersion of these excitations.

\begin{figure}[h]
\begin{center}
\includegraphics[width=7cm]{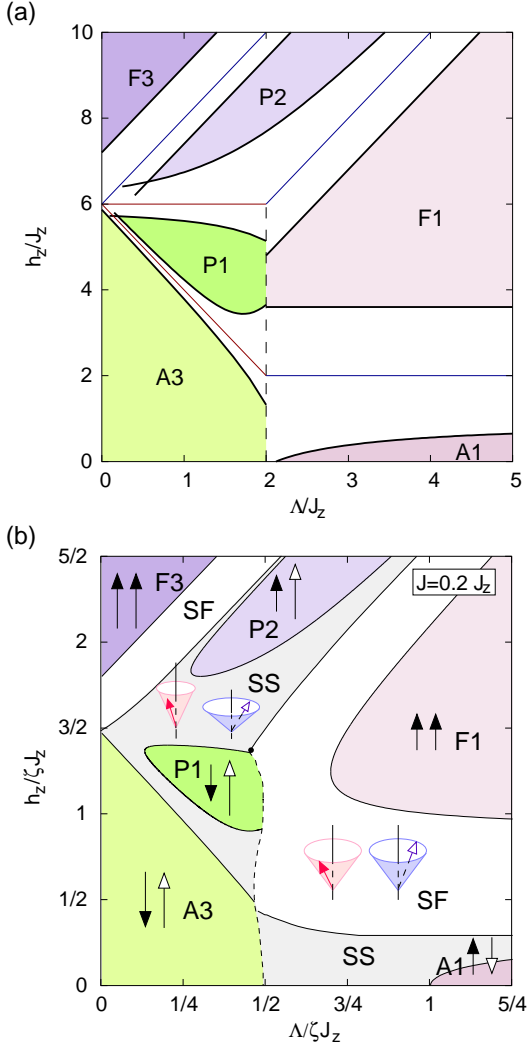}
\caption{(color online) (a) Instabilities (thick lines) of the gapped phases in the square lattice as obtained from the  perturbation theory for $J/J_z=0.2$. We also show the $J=0$ phase boundaries of Fig.~\ref{fig:ising} with thin lines for comparison. (b) Variational phase diagram as  function of $\Lambda/\zeta J$ and $h_z/\zeta J$ for large exchange anisotropy $J/J_z=0.2$ and a bipartite lattice with $\zeta$ neighbors. Dashed lines stand for first order, while solid lines denote second order phase transitions. The solid dot ending the first order transition represent a tricritical point.}
\label{fig:ising2}
\end{center}
\end{figure}

In the case of a uniform order the spins on the two sublattices are equal, and the perturbational expansion of the excitation energy is simple.
Let us pick an example, e.g. the instability of the fully polarized phase $F3$ towards the plateau $P2$. In $F3$ the ground state is $\prod_{j}|\Uparrow_j\rangle$. A spin excitation in this case corresponds to lowering the $\Uparrow$ spin to a $\uparrow$ on a given site, with a diagonal energy cost
\begin{equation}
\Delta E= h -2\Lambda -\frac{3}{2}\zeta J_z.
\end{equation}
The off--diagonal terms move the excitations onto the neighboring sites, as shown in Fig. \ref{fig:Hoppings}(a), with a 
\begin{equation}
  \langle\uparrow_i\Uparrow_j|\mathcal{H}|\Uparrow_i\uparrow_j\rangle=\frac{3J}{2}\;
\end{equation}
hopping amplitude, leading to the following dispersion:
\begin{equation}
\omega_{{\bf k}} = h-2\Lambda +\frac{3}{2}\zeta \left(J \gamma_{\mathbf{k}} - J_z\right) .
\label{eq:omega1magnonF3}
\end{equation}
Here 
\begin{equation}
 \gamma_{\mathbf{k}} = \frac{1}{\zeta}\sum_{\bm{\delta}} \exp(i \mathbf{k}\cdot\bm{\delta}),
 \label{eq:gammak}
\end{equation}
where the summation is over the vectors $\bm{\delta}$ pointing toward the $\zeta$ nearest neighbors. The quantity $\gamma_{\mathbf{k}}$ takes its minimal value $-1$ at $\mathbf{k} = (\pi,\dots)$, and its maximal value $1$ at $\mathbf{k} = (0,\dots)$. For the one--dimensional chain ($\zeta = 2$)
\begin{equation} 
 \gamma_{\mathbf{k}} = \cos k_x  ,
\end{equation}
and 
\begin{equation} 
 \gamma_{\mathbf{k}} = \frac{1}{2}\left(\cos k_x +\cos k_y \right) 
\end{equation}
for the square lattice ($\zeta = 4$).
In the $F_3$ phase this excitation is gapped with a minimum at $\mathbf{k}=(\pi,\dots)$, and lowering the magnetic field the gap closes when
\begin{eqnarray}
 h_{\text{sat}} = \frac{3}{2}\zeta \left(J_z+J\right)+2\Lambda .
 \label{eq:hsat}
\end{eqnarray}
Instabilities of this kind are summarized in Eqs.~(\ref{eq:wF1P2})-(\ref{eq:wF3P2}), the corresponding critical fields are shown in Table~\ref{tab:1st_ord_instab}, and are plotted in Fig.~\ref{fig:ising2}(a) for $J/J_z=0.2$. We shall mention that these results are not independent from the mapping we discussed in the previous subsection.

We note that in the case of the $F_3$ phase Eqs.~(\ref{eq:omega1magnonF3}) and (\ref{eq:hsat}) are exact, while for $F1$ higher order terms in $J/J_z$ appear in the dispersion. 

\begin{table}
\caption{(color online) Summary of instabilities of uniform phases.%
\label{tab:1st_ord_instab}
}
\begin{center}
\begin{ruledtabular}
\begin{tabular}{lccc}
\raisebox{0.5ex}[0pt]{} &
\raisebox{0.5ex}[0pt]{$\Delta E$} & 
\raisebox{0.5ex}[0pt]{hopping amplitudes} & 
\raisebox{0.5ex}[0pt]{$h_c$} \\ 
\hline
$F3\to P2$ & $h-2\Lambda-6 J_z$ & $3 J/2$ & $2\Lambda + 6 J_z+ 6 J$\\
$F1\to P2$ & $2 J_z-h+2\Lambda$ & $6 J$ & $2\Lambda + 2J_z - 6 J$\\
$F1\to A1$ & $-2 J_z+h$ & $2 J$ & $2J_z+ 8 J$\\
\end{tabular}
\end{ruledtabular}
\end{center}
\end{table}

\subsection{Dispersion of spin--excitations in translational symmetry breaking states on the square lattice} 
\label{sec:ising_2sublattice_pert}

The instability (softening) of the excitations in the two-sublattice gapped phases ($A1$, $A3$, $P1$, and $P2$) that break the translational symmetry occur in the second order of exchange coupling $J$. Namely, the on--site excitations on the two sublattices have different energy, and depending on the energy difference we shall apply a different scheme for the degenerate perturbation calculation. As an example, we discuss the lower instability of the 2/3--plateau $P2$ phase. 

The  wave function in the Ising limit of the $P2$ phase is given by 
\begin{equation}
 |\Psi^{P2} \rangle = \prod_{j\in A} \prod_{j' \in B} |\uparrow_j \rangle  |\Uparrow_{j'}\rangle .
\end{equation}
 Applying the $S_i^-$ operator on the $A$ and the $B$ sublattice, we get 
 \begin{eqnarray} 
  |\Phi^{A}_{i}\rangle &=& |\downarrow_i\rangle \prod_{j \in A \atop j\neq i} \prod_{j'\in B} |\uparrow_j   \rangle  |\Uparrow_{j'}\rangle, \\
  |\Phi^{B}_{i}\rangle &=& |\uparrow_i\rangle \prod_{j \in A} \prod_{j'\in B\atop j'\neq i} |\uparrow_j   \rangle  |\Uparrow_{j'}\rangle,
\end{eqnarray} 
with diagonal excitation energies  
\begin{eqnarray}
  \Delta E_A &=& h - 6 J_z , \\
  \Delta E_B &=& h - 2 \Lambda - 2 J_z ,
\end{eqnarray}
respectively.
The two energies are identical when $\Lambda = 2 J_z$ --- this is actually the phase boundary between the $P_1$ and $F_1$ phase in the Ising phase diagram, Fig.~\ref{fig:ising}. 

\begin{figure}[ht]
\centering
\includegraphics[width=7cm]{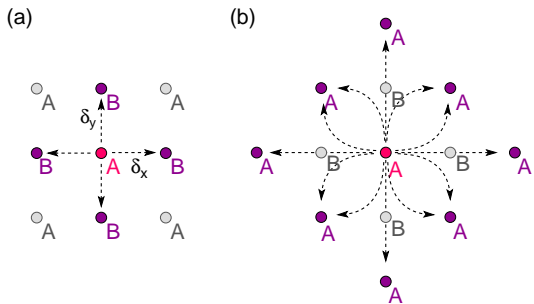}
\caption{(color online) (a) Schematic figure for the first order hopping process that occurs during the instability of uniform phases $F1$ and $F3$, where the dispersion is $\propto 4 \gamma_{\mathbf{k}}$. (b) Schematic representation of the second neighbor correlated hopping that gives the dipersion $\propto 16 \gamma_{\mathbf{k}}^2$. There are $8$ neighboring places where the magnon can hop through a virtual state on the $B$-site. \label{fig:Hoppings}}
\end{figure}

First we discuss the case when the energy difference is larger than $J$: when $\Delta E_B - \Delta E_A = 4 J_z - 2 \Lambda \gg J$, the ground state manifold is given by the $|\Phi^{A}_{i}\rangle$ states. 
Since 
$  \langle\Phi^{B}_{i}| \mathcal{H} |\Phi^{A}_{i'}\rangle = \sqrt{3} J $ for neighboring $i$ and $i'$ sites and $\langle\Phi^{A}_{i}| \mathcal{H} |\Phi^{A}_{i'}\rangle = 0$,  the $|\downarrow\rangle$ excitation acquires dispersion in a second order process in $J$, where the $|\uparrow\rangle$ excitations  on the $B$ sublattice can be viewed as virtual state [see Fig.~\ref{fig:Hoppings}(b)]. This leads to 
\begin{equation}
\omega_{P2 \rightarrow P1}(\mathbf{k}) = h - 6 J_z  - \frac{3 J^2}{4 J_z - 2\Lambda } 16 \gamma^2_{\mathbf{k}} + \omega_{P2 \rightarrow P1}^{(2)}
\label{eq:wP2P1simpl}
\end{equation}
where the $\omega_{P2 \rightarrow P1}^{(2)}$ denotes additional second order contributions in $J$ that are independent of $\mathbf{k}$ --- the full form of the dispersion is given in Eq.~
(\ref{eq:wP2P1}). In other words, the gap closes quadratically for small values of $J$. A similar calculation can be done for the $\Delta E_A - \Delta E_B = 2 \Lambda  - 4 J_z \gg J$ case, when the 
ground state manifold is given by the $|\Phi^{B}_{i}\rangle$ states, and we  similarly get a  dispersion, Eq.~(\ref{eq:wP2F1}) in the Appendix, where the hopping amplitude is quadratic in $J$ (we note that an additional virtual state assists the hopping).

When the two excitation have equal energy at $\Lambda = 2 J_z$, the perturbation theory outlined above obviously fails [the hopping amplitudes in both Eqs.~(\ref{eq:wP2F1}) and (\ref{eq:wP2P1}) diverge]. In that case we shall include both $|\Phi^{A}_{i}\rangle$ and $|\Phi^{B}_{i}\rangle$ states into the ground state manifold. Actually, we can do it also when the energies are not equal, and to get the dispersion of the spin excitations, we need to diagonalize the following $2\times 2$ problem in  $\mathbf{k}$ space:
\begin{equation}
\mathcal{H'}_{P2} = 
\left(
\begin{array}{cc}
h - 6 J_z
&  
4 \sqrt{3} J \gamma_{\mathbf{k}}   
\\
4 \sqrt{3} J \gamma_{\mathbf{k}}   
&     
h - 2 J_z - 2 \Lambda \end{array}
\right) 
\end{equation}
where we neglected second order contributions. The $2\times 2$ matrix is easily diagonalized, leading to the 
\begin{equation}
\omega_{\mathbf{k}}= h-4 J_z - \Lambda \pm\sqrt{\left(\Lambda-2J_z\right)^2+ 48 J^2 \gamma_{\mathbf{k}}^2}
\end{equation}
dispersion. We notice that for $\Lambda = 2 J_z$ the dispersion becomes linear in $J$, while for $J \ll |\Lambda-J_z|$ expanding the square root we get 
\begin{equation}
\omega_{\mathbf{k}}= h-4 J_z - \Lambda \pm \left(\Lambda-2J_z\right) 
\pm \frac{24 J^2 \gamma_{\mathbf{k}}^2}{\Lambda-2J_z}.
\end{equation}
In other words, we obtain the result of the second order perturbation theory, Eq.~(\ref{eq:wP2P1simpl}). To be consistent, we shall take into account all the second order processes that contribute to the dispersion. This can be done systematically, and the full expression is given in Eq.~(\ref{eq:H2x2P2}).
The critical field at which the gap vanishes can then be determined without difficulty, and the instabilities of this type, given by Eqs.~(\ref{eq:H2x2A1}), (\ref{eq:H2x2P1}), and (\ref{eq:H2x2P2}) are shown in Fig.~\ref{fig:ising2}(a) for  $J/J_z=0.2$.

\section{Variational Phase Diagram}\label{sec:variational}

In this section we construct the phase diagram variationally, assuming either uniform or two--sublattice ordering.
We search for the ground state in the following site--factorized variational form: 
\begin{equation}
 |\Psi\rangle = \prod_{i\in A}\prod_{j\in B} |\psi_{A}\rangle_i |\psi_B\rangle_j \;,
\label{eqn:var-ansatz}
\end{equation}
where
\begin{equation}
| \psi_A \rangle \propto u_{0} |\Uparrow\rangle+e^{i\xi_1}u_{1}|\uparrow\rangle+e^{i\xi_2}u_2|\downarrow\rangle+e^{i\xi_2}u_3|\Downarrow\rangle 
\label{eq:varpsiABB}
\end{equation}
and a similar expression for $| \psi_B \rangle$. 
In the general case, there are 6 independent  variational parameters for 
$| \psi_A \rangle$ and another 6 for $| \psi_B \rangle$ that are determined by minimizing the ground state energy
\begin{equation}
  E = \frac{\langle \Psi | \mathcal{H} | \Psi \rangle}{\langle \Psi | \Psi \rangle} \;.
  \label{eqn:var-energy}
\end{equation}
Recalling that the Hamiltonian is $O(2)$ symmetric and commutes with the $\hat S^z = \sum_i S^z_i $ operator, the state rotated by $\varphi$ around the $z$ axis and given by the $\exp(- \varphi \hat S^z) |\Psi\rangle$ wave function has the same energy as the state described by $|\Psi\rangle$. We can therefore reduce the number of independent parameters for sites $A$ from 6 to 5, so in total we have reduced the number of independent parameters from 12 to 11. It appears, however, that all the phases we have found are coplanar, and after a suitable rotation the amplitudes in the wave function can all be chosen to be real.

 The site--factorized variational wave function (\ref{eqn:var-ansatz}) is actually indifferent to the connectivity of the lattice, the only information about the lattice that enters the expression of the energy is the number of the neighbours $\zeta$. For concreteness, we look at the case of the square lattice, however the results can be easily generalized to any bipartite lattice by replacing $J \rightarrow \zeta J/4$ and $J_z \rightarrow \zeta J_z/4$ in equations and phase diagrams shown below.

 Before proceeding to discussion of the phase diagrams, let us mention briefly that for the gapped phases the variational wave function is of the same form as it is in the Ising--limit, when we neglected the off--diagonal terms. Similarly, the expressions for the ground state energy are also identical, since to get a contribution from the off-diagonal 
$\hat S^x_i \hat S^x_j+\hat S^y_i \hat S^y_j$ term, we need to tilt the spins out of the $z$ axis. 
Furthermore, the boundary of the gapped phases, assuming a continuous phase transition, can be determined by studying the stability of the gapped variational wave  function $|\Psi_0\rangle$: the 0  eigenvalue of the $\partial^2 E/\partial u_\alpha \partial u_\beta$ indicates a second order phases transition, where $u_\alpha$ and $u_\beta$ are coefficients of a wave functions that is orthogonal to $|\Psi_0\rangle$.


\subsection{Phase diagram in zero magnetic field}

\begin{figure}[h!]
\begin{center}
\includegraphics[width=7cm]{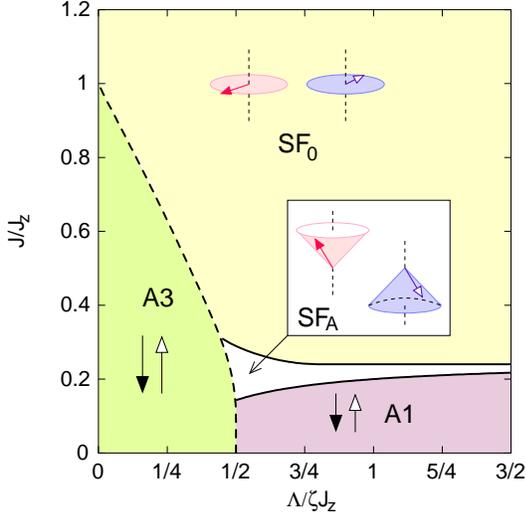}
\caption{(color online) Variational phase diagram for $h=0$ as the function of $\Lambda/J_z$ and $J/J_z$. Solid lines stand for continuous (second order) phase boundaries, while the dashed line denotes the first order phase boundary of the phase $A3$ .}

\label{fig:h0_pd}
\end{center}
\end{figure}

First let us take a look at the zero field phase diagram. We find two -- the completely and the partially aligned -- axial antiferromagnetic states, $A3$ and $A1$ as well as a superfluid $U(1)$ phase between them. This latter is referred to as a planar state in Ref.~[\onlinecite{Solyom1984}] for the spins are aligned in the lattice plane, but also can be called superfluid since spin-rotation symmetry breaking phases exhibit finite spin stiffness\cite{Seabra2011} that is the property of such phases. In the following we refer to this phase as $SF_0$. Between the planar superfluid $SF_0$ and the two axial antiferromagnets $A1$ and $A3$ an additional superfluid phase appears. The in-plane components of this conical antiferromagnet have the same properties as the planar superfluid but it exhibits finite staggered magnetization too, inheriting the property of the antiferromagnetic phases. Therefore we call this phase $SF_A$. A schematic figure of the various phases is shown in the phase diagram in Fig. \ref{fig:h0_pd}.

\begin{figure}[h!]
\begin{center}
\includegraphics[width=7cm]{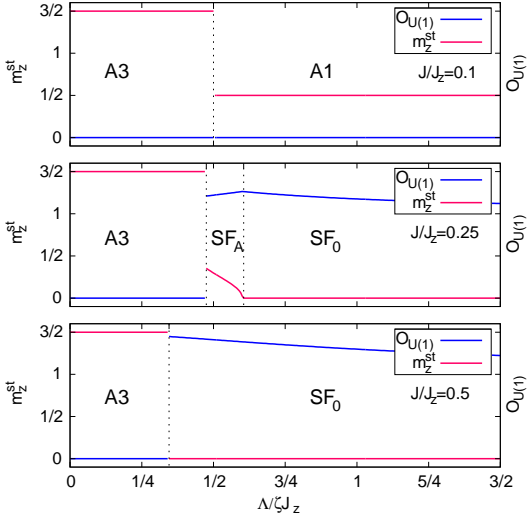}
\caption{(color online) Order parameters as the function of $\Lambda/J_z$ for different values of $J/J_z$. The axial antiferromagnets have finite staggered magnetization, and zero $U(1)$ order parameter. In the planar superfluid phases $m^{st}_{z}=0$ but the expectation value of $O_{U(1)}$ is finite, while the conical antiferromagnet exhibits both type of order.}
\label{fig:h0_ordP}
\end{center}
\end{figure}
The relevant order parameters for zero external field are the staggered magnetization $m^{\text{st}}_{z}=\frac{1}{2}| S^{z}_A- S^{z}_B|$, and the superfluid order parameter $O_{U(1)}=\frac{1}{2}|{\bf S}^{\bot}_A-{\bf S}^{\bot}_B|$, where ${\bf S}^{\bot}_j=(S^x_j,S^y_j)$. $O_{U(1)}$ is actually the in-plane staggered magnetization. The expectation values of order parameters as a function of $\Lambda/J_z$ are shown in Fig. \ref{fig:h0_ordP} for various values of $J/J_z$.

The first order phase boundary between the two axially aligned antiferromagnetic phases is $\Lambda=2 J_z$. It can be determined by comparing the ground state energies of  $A3$ and $A1$ listed in Table \ref{tab:ising_wavef_orderp}. 


The ground state wave functions of sites $A$ and $B$ in the planar superfluid phase $SF_0$ can be expressed as
\begin{eqnarray}  
|\Psi_A\rangle &=& e^{-i \varphi \hat S^z_A} |\Psi_{\text{SF}}\rangle, \\
|\Psi_B\rangle &=& e^{-i (\varphi+\pi) \hat S^z_B} |\Psi_{\text{SF}}\rangle,
\end{eqnarray}
with a single variational parameter $\eta$:
\begin{equation}
|\Psi_{\text{SF}}\rangle = \frac{1}{\sqrt{3 \eta^2 + 1}}\left( |\Uparrow\rangle + \sqrt{3}\eta |\uparrow\rangle+ \sqrt{3}\eta |\downarrow\rangle+  |\Downarrow\rangle \right) .
\label{eq:SF0_grst}
\end{equation}
$\varphi$ can take arbitrary value and is related to the $U(1)$ symmetry breaking (we recall that the Hamiltonian commutes with the $\hat S^z$ operator), as it determines the direction that the spins point to in the $xy$ plane: 
\begin{eqnarray}
\langle\Psi_A|\hat S^x_A|\Psi_A\rangle &=& \frac{3 \eta  (\eta +1)}{3 \eta ^2+1} \cos \varphi ,\\
\langle\Psi_A|\hat S^y_A|\Psi_A\rangle &=& \frac{3 \eta  (\eta +1)}{3 \eta ^2+1} \sin\varphi ,\\
\langle\Psi_A|\hat S^z_A|\Psi_A\rangle &=& 0 .
\end{eqnarray}
The ground state energy as the function of parameter $\eta$ reads
\begin{eqnarray}
\frac{E^{SF_0}_{0}(\eta)}{N} = \frac{3}{4} \frac{\eta^2 + 3}{3 \eta^2 + 1} \Lambda -\frac{18 \eta^2
   \left( \eta + 1 \right)^2}{\left( 3\eta^2 + 1 \right)^2} J .
  \end{eqnarray}
In the energy expression the $J_z$ term is absent, as this wave function has only spin component in the $xy$ plane. 
Minimizing the energy gives a cubic equation for $\eta$. However, a given $\eta$ value minimizes the energy for the 
\begin{equation}
\frac{\Lambda}{J} = \frac{3 (\eta^2 -1) (3 \eta +1)}{3 \eta ^2+1}
\end{equation}
parameter in the Hamiltonian.
For small values of $\Lambda$ we find
\begin{eqnarray}
\eta=1+\frac{\Lambda }{6 J}+\frac{\Lambda ^2}{144 J^2}+O\left(\Lambda ^3\right)
\end{eqnarray}
and the ground state energy can be approximated as:
\begin{eqnarray}
E^{SF_0}_{0}=-\frac{9}{2} J +\frac{3}{4}\Lambda-\frac{\Lambda^2}{16 J}+O(\Lambda^3) \;
\end{eqnarray}
that gives the phase boundary with the antiferromagnetic phase $A3$
\begin{eqnarray}
J=J_z-\frac{\Lambda}{3}-\frac{\Lambda^2}{72 J_z}+ O(\Lambda^3)\;
\end{eqnarray}
as seen in Fig.~\ref{fig:h0_pd}.

For $\Lambda =0$, when the anisotropy is absent, $\eta=1$ and Eq.~(\ref{eq:SF0_grst}) is just a spin coherent state of the spin--3/2 N\'eel-state of the SU(2) symmetric Heisenberg model rotated into the $xy$ plane. For $\Lambda >0$ the $S^z=\pm 3/2$ components in the wave function are suppressed. In the $\Lambda \rightarrow +\infty$ limit the $\eta = \Lambda/3+O(1)$ and we are left with a wave function with $S^z=\pm 1/2$ spin components only. Out of these two states we can mix a spin pointing to arbitrary direction, however the length of the spin is not constant -- it is the largest when lying in the $xy$ plane (the length is then 1 as opposed to 1/2 when pointing in the $z$ direction, a consequence that they are still $S=3/2$ spin). For this reason the antiferromagnetic exchange term gains the most energy with the planar spins, as in Eq.~(\ref{eq:SF0_grst}). When the exchange interaction becomes anisotropic, and the $\hat S^z_i\hat S^z_j$ term becomes strong, this energy can compensate the directional length dependence of the spin, and can choose a spin configuration with a finite $z$ and $xy$ component. This happens in the conical superfluid phase, denoted by $SF_A$ in Fig.~\ref{fig:h0_pd}.

The phase boundary of the conical superfluid phase ($SF_A$) towards the planar phase ($SF_0$) and fully polarized AFM phase ($A3$) is a complicated expression. It is shown in Fig. \ref{fig:h0_pd}. Starting from phase $A1$ at a given $\Lambda$ value, a second order phase transition occurs to the superfluid phase $SF_A$. When the exchange coupling $J$ is large enough, in-plane spin components appear continuously as we reach into $SF_A$.
The ground state can be expressed as it follows:
\begin{eqnarray}  
|\Psi_A\rangle &\propto& e^{-i \varphi \hat S^z_A} 
 \left( |\Uparrow\rangle + u |\uparrow\rangle+ v |\downarrow\rangle+ w |\Downarrow\rangle \right) \\
|\Psi_B\rangle &\propto& e^{-i (\varphi+\pi) \hat S^z_B} 
\left( w |\Uparrow\rangle +  v |\uparrow\rangle + u |\downarrow\rangle + |\Downarrow\rangle \right)\;
\label{eq:SFA_grst}
\end{eqnarray}
with real $u$, $v$ and $w$ variational parameters. 

The instability of the partially aligned AFM phase $A1$ against canting gives the phase boundary 
\begin{eqnarray}
J=\frac{J_z (J_z-\Lambda)}{J_z-4\Lambda} \;
\end{eqnarray}
between $A1$ and $SF_A$.

The same model for one dimension has been treated by mean field calculations in Ref.~[\onlinecite{Solyom1984}] for quantum spin $1/2$, $1$ and $3/2$. The phase diagram for the case $S=3/2$ is similar to our findings, however the conical superfluid phase $SF_A$ is missing due to a more restricted variational wave function they used.


\subsection{Heisenberg exchange with on--site anisotropy}

\begin{figure}[h!]
\begin{center}
\includegraphics[width=7cm]{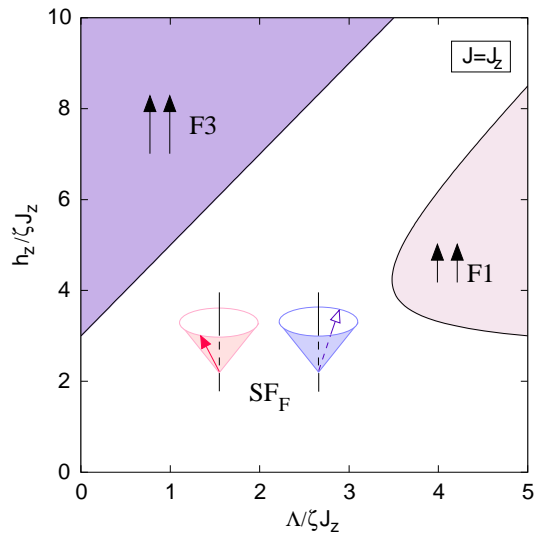}
\caption{(color online) Phase diagram in the function of $\Lambda/J$ and $h_z/J$ ($J_z=J$).}
\label{fig:hz_L_pd}
\end{center}
\end{figure}

In the following we discuss the phase diagram as the function of magnetic field and single-ion anisotropy when the exchange between two neighboring sites is SU(2) symmetric (i.e. the $J=J_z$ Heisenberg model with on--site anisotropy). The phase diagram outlined in Fig.~\ref{fig:hz_L_pd} was calculated by the variational method introduced above.

\begin{figure}[h!]
\begin{center}
\includegraphics[width=7cm]{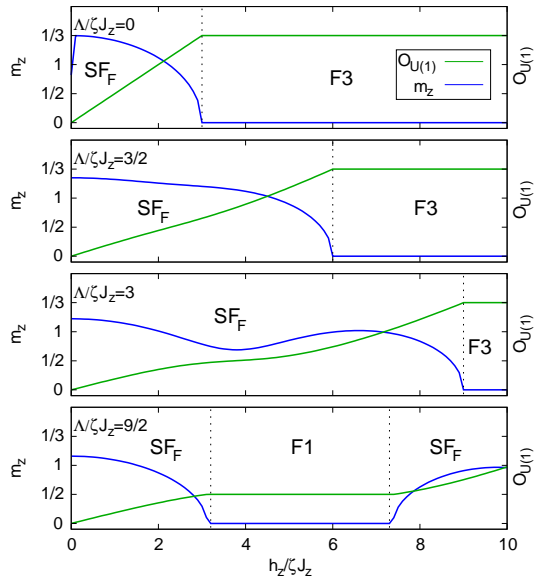}
\caption{(color online) Order parameters as the function of magnetic filed for different values of $\Lambda$ parameter. The fully and partially polarized ferromagnets $F3$ and $F1$ exhibit finite magnetization $m_z$, while in the conical ferromagnetic phase $SF_F$ the expectation values of both $m_z$ and $O_{U(1)}$ are finite.}
\label{fig:JeqJz_ordP}
\end{center}
\end{figure}

On the $h_z=0$ line the ground state is the planar superfluid phase ($SF_0$) introduced previously. As the magnetic field becomes finite, the spins cant out of the plane continuously, and the superfluid ground state $SF_F$ exhibits finite magnetization $m_z$ alongside the finite staggered in-plane order parameter $O_{U(1)}$ [Fig.~\ref{fig:JeqJz_ordP}]. A schematic figure of the conical $SF_F$ is shown in Fig. \ref{fig:hz_L_pd} and the ground state wave function can be characterized as:
\begin{eqnarray}  
|\Psi_A\rangle &\propto& e^{-i \varphi \hat S^z_A} 
 \left( |\Uparrow\rangle + u |\uparrow\rangle+ v |\downarrow\rangle+ w |\Downarrow\rangle \right) \\
|\Psi_B\rangle &\propto& e^{-i (\varphi+\pi) \hat S^z_B} 
\left( |\Uparrow\rangle + u |\uparrow\rangle+ v |\downarrow\rangle+ w |\Downarrow\rangle \right)\;
\label{eq:SFF_grst}
\end{eqnarray}
where $u$, $v$, and $w$ are all real numbers.

The ground state energies of the axial ferromagnetic phases and the fully polarized ferromagnetic state are given in Table~\ref{tab:ising_wavef_orderp}, 

Analytical expression for the ground state energy of $SF_F$ is beyond our reach, however, the phase boundary with the neighboring $F3$ phase can be given by calculating the critical field for the polarized phase. This is exactly the same as the instability approximation for $F3$ in the case of the Ising limit, and is given by the same Eq.~(\ref{eq:hsat}). Above the saturation field the fully polarized ferromagnetic phase is stabilized. 
For large enough values of $\Lambda$ the spins become shorter and the partially polarized plateau phase $F1$ emerges. 
Calculating the instability of  $F1$, the phase boundary  turns out be
\begin{eqnarray}
 h = J+2 J_z+\Lambda \pm \sqrt{J^2-14J\Lambda+\Lambda^2}.
\label{eq:bound_f1}
\end{eqnarray}

As expected, from the mapping to the effective XXZ model (Sec.~\ref{sec:mappingXXZ}), we found no evidence for gapped phases that break the translational symmetry.


\subsection{The effect of exchange anisotropy and the emergence of supersolid phase.}\label{sec:supersolid}

Finally let us examine the collective effect of exchange and single-ion anisotropies as well as the magnetic field.
In the previous subsection we learned that only the ferroaligned spins in $F1$ and $F3$ are present as gapped phases for the case of the Heisenberg exchange ($J_z=J$), with a superfluid phase (canted antiferromagnet)  in between them. As the value of $J/J_z$ is lowered, islands of plateaus and antiferromagnetic phases emerge in the sea of the superfluid phase.  We choose a relatively large anisotropy $J/J_z=0.2$, as in that case we learned from the perturbational expressions that the 2--fold degenerate gapped phases might be stable, as shown in Fig.~\ref{fig:ising2}(a). Indeed, the variational phase diagram, shown in Fig. \ref{fig:ising2}(b), displays all the phases we were looking for: The superfluid phase takes place around the axial ferromagnets, 
while between the plateaus and axial antiferromagnetic phases -- i.e. the gapped phases that exhibit staggered diagonal magnetic order -- a very robust supersolid phase arises. 

The extension of the supersolid around the phases $A1$ and $P2$ is the broadest at their tips, when $\Lambda$ is not too large. As we increase $\Lambda$, the supersolid region decreases, and eventually vanishes for $\Lambda\rightarrow +\infty$. Since in this limit the mapping to the XXZ model becomes exact (Sec.~\ref{sec:mappingXXZ}), our finding is also consistent with numerical works on the XXZ model on the square lattice that do not seem to find supersolid.\cite{Masanori1997,Seiji2002,Batrouni2000}

\begin{figure}[h!]
\begin{center}
\includegraphics[width=8cm]{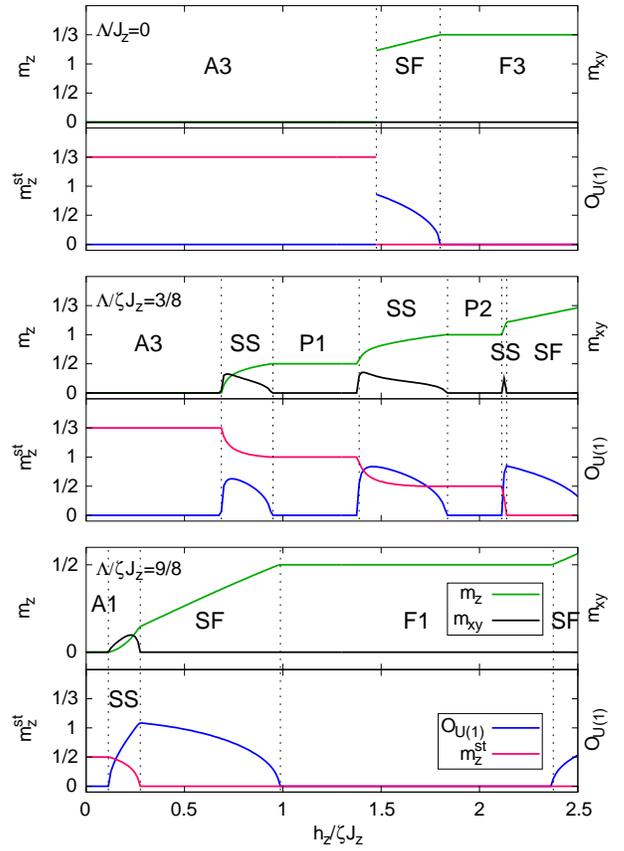}
\caption{(color online) Expectation value of order parameters per site as the function of $h/J_z$ for different values of $\Lambda/J_z$. In the axial antiferromagnetic phases $A1$ and $A3$ only the staggered magnetization has finite expectation value. For $\Lambda=0$ there is a first order phase transition from the completely polarized $A3$ phase to the superfluid phase. All the other field induced transitions are second order transitions. The superfluid phase exhibits finite magnetization $m_z$ and finite staggered in-plane magnetization $O_{U(1)}$. The ferromagnetically ordered  phases $F3$ and $F1$ are characterized by finite magnetization $m_z$, while the plateaus ($P1$ and $P2$) have an additional finite staggered magnetization $m^{st}_z$. In the supersolid phase all four order parameters have finite expectation values.}
\label{fig:Jz5_ordP}
\end{center}
\end{figure}

The variational calculation finds all the phase boundaries to be of second order, except a single first order one around $\Lambda \approx 2 J_z$ [shown in dashed line in Fig.~\ref{fig:ising2}(b)] that is 
inherited from the $J=0$ phase diagram, Fig.~\ref{fig:ising}. 

The expression of the phase boundaries of the axial ferromagnetic phases are the same as in the Heisenberg limit (see Eqs.~(\ref{eq:hsat}) and (\ref{eq:bound_f1})). We determined the phase boundaries of the plateaus and axial antiferromagnetic states by calculating spin wave instability. We found that the boundary for the $2/3$ plateau can be given as
\begin{eqnarray}
\Lambda=\frac{h}{2}-\Theta\pm\sqrt{(\Theta-3J_z)^2-9J^2},
\end{eqnarray}
with $\Theta=2 J_z+\frac{6J^2}{h/2-3 J_z}$. Similar calculations give
\begin{eqnarray}
h&=&\sqrt{2(J^2_z-J^2)+2 (J_z-\Lambda)^2-2\Sigma},\nonumber\\
\Sigma&=&\sqrt{(J^2-\Lambda(\Lambda-2J_z))^2+32 J^2\Lambda(2\Lambda-J_z)},
\end{eqnarray}
for the phase boundary of the partially polarized axial antiferromagnetic phase $A1$,
\begin{eqnarray}
\Lambda=\Theta-\frac{h}{2}\pm\sqrt{(\Theta-3J_z)^2-9J^2}\;
\label{eq:bound_P1}
\end{eqnarray}
with $\Theta=2 J_z+\frac{h}{2}+\frac{6J^2}{h/2-3 J_z}$ for the phase boundary of $P1$ plateau, and
\begin{eqnarray}
\Lambda=3 J_z-\sqrt{\frac{h^2}{4}+9 J^2}\;
\label{eq:bound_A3}
\end{eqnarray}
for the boundary of the axial antiferromagnetic phase $A3$. When $J=0$ Eq. (\ref{eq:bound_P1}) and (\ref{eq:bound_A3}) give back the $h=6J_z-2\Lambda$ phase boundary that separates $A3$ and $P1$ in the Ising limit.
The ground state energies and phase boundaries for the superfluid and supersolid phases can only be obtained numerically. The ground state wave function for  the superfluid with ferromagnetic $m^z$ is given by Eq. (\ref{eq:SFF_grst}) , and for the supersolid by
\begin{eqnarray}  
|\Psi_A\rangle &\propto& e^{-i \varphi \hat S^z_A} 
 \left( |\Uparrow\rangle + u |\uparrow\rangle + v |\downarrow\rangle + w |\Downarrow\rangle \right) \\
|\Psi_B\rangle &\propto& e^{-i (\varphi+\pi) \hat S^z_B} 
\left( |\Uparrow\rangle + u' |\uparrow\rangle + v' |\downarrow\rangle + w' |\Downarrow\rangle \right),\;
\label{eq:SS_grst}
\end{eqnarray}
where $u$, $u'$, $v$, $v'$, $w$, and $w'$ are all real. 
Fig.~\ref{fig:Jz5_ordP} shows the evolution of the order parameters which can be used to find out the nature of the phases as we increase the magnetic field for a few selected values of $\Lambda/J_z$.

\subsection{S=1 phase diagram}

\begin{figure}[bth]
\includegraphics[width=6cm]{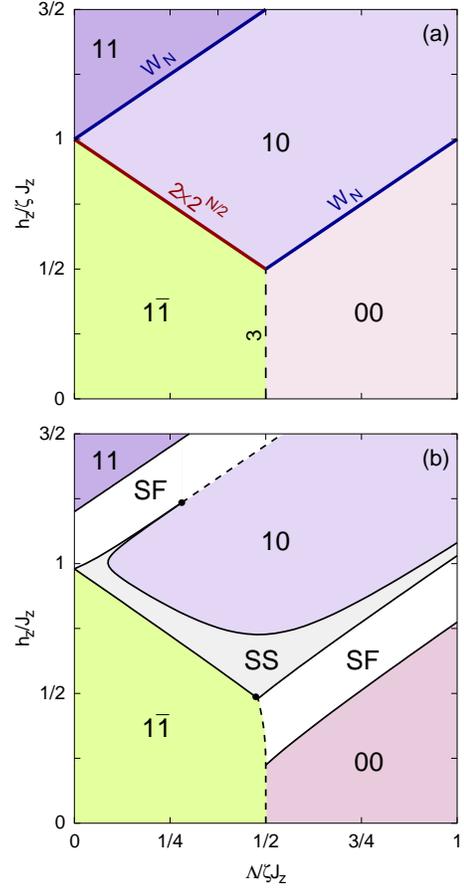}
\caption{(color online) The phase diagram of the anisotropic S=1 model in the (a) Ising--limit for a bipartite lattice with coordination number $\zeta$ and (b) for the square lattice ($\zeta=4$) when $J = 0.2 J_z$, obtained from the variational calculation. }
\label{fig:PD_S=1}
\end{figure}

At this stage, it is useful to compare the predictions of the variational approach to a better studied problem. Supersolid phases have been found in spin--1 anisotropic Heisenberg antiferromagnet in Ref.~\onlinecite{Sengupta2007}, so we constructed the variational phase diagram for this model as well. The Hamiltonian is identical to Eq.~\ref{eq:Hamiltonian}, but now with $S=1$ spin operators. The phase diagrams, for vanishing $J$ and $J=0.2 J_z$, are shown in Fig.~\ref{fig:PD_S=1}. In the Ising limit,  Fig.~\ref{fig:PD_S=1}(a), we find two uniform phases (denoted as $00$ and $11$, using the values of the $S^z$ components) and two phases breaking the translational symmetry: the $1\bar1$ with zero magnetization and the $10$ one--half magnetization plateau. The XXZ--like physics can be identified for the transition between the $11$,$10$, and $00$ phases, where the supersolidity is a fragile phase. The region between the $10$ and $1\bar1$ is of different nature, and we expect the supersolid to be robust in this part of the phase diagram. And that is exactly the region where Ref.~\onlinecite{Sengupta2007} found supersolidity. Furthermore, the nature of the phase transitions is also in qualitative agreement, inasmuch the order of the phase transitions is concerned. Specifically, we recover the first order transition between the upper boundary of the $10$ phase and the superfluid. It is also useful to compare Fig.~\ref{fig:PD_S=1}(b) to the phase diagram of the one--dimensional chain obtained by DMRG:\cite{Peters2010} 
the extent of the gapped phases is reduced in the chain, and the supersolid survived only in a small region close to the $1\bar1$ phase.

The calculation of the phase diagram is quite straightforward -- assuming two sublattice order, the variational wave function is given by Eq.~(\ref{eqn:var-ansatz}), now with
\begin{equation}
| \psi_A \rangle \propto u_{1} |1\rangle + e^{i\xi_0}u_{0}|0\rangle +e^{i\xi_{\bar 1}}u_{\bar 1}|\bar 1\rangle
\label{eq:varpsiABB}
\end{equation}
and a similar expression for $| \psi_B \rangle$. 
We are now dealing with 8 independent  variational parameters altogether, that can be reduced to 7 by using the $U(1)$ symmetry of the model. Similarly to the spin--3/2 case, we get solutions where all the amplitudes can be chosen to be real numbers for the Hamiltonian we look at.

The saturation field is given by 
$h_{\text{sat}} = \Lambda + \zeta J_z + \zeta J$, and from the stability analysis of the  $1\bar1$, $00$, and $10$ gapped phases we get the following equations for their phase boundaries
\begin{eqnarray}
&& h^2 = ( \zeta J_z - \Lambda - \zeta J ) (  \zeta J_z - \Lambda  + \zeta J ),\\
&& h^2 = \Lambda ( \Lambda - 2\zeta J ), \\
&& (h-\Lambda ) (\zeta J_z + \Lambda -h) (h-\zeta J_z+\Lambda ) = 2\zeta^2 J^2 \Lambda, 
\end{eqnarray}
respectively.

\section{Supersolid in the one-dimensional model}
\label{sec:chain}
In this section we complement the variational study using a variant of the Density Matrix Renormalization Group\cite{White1992} (DMRG) method on the anisotropic $S=3/2$ spin chain.  A Quantum Monte-Carlo study has shown that a supersolid phase can realized in the anisotropic $S=1$ spin chain,\cite{Sengupta2007/2} a result confirmed by DMRG calculations in Refs.~\onlinecite{Peters2009,Peters2010,Rossini2011}. Therefore it is plausible that a supersolid states is also present in the anisotropic spin--3/2 chain. 

We map out the phase diagram  for the chain and  search for  signatures of supersolid phases. The DMRG method we used optimizes variationally a wavefunction based on a matrix-product state\cite{Oestlund1995} (MPS) Ansatz for an infinite chain.  Algorithms using this approach are efficient in one dimensional systems because they exploit the fact that the ground-state wave functions are only slightly entangled. For mapping out the phase diagram, we used comparably small matrix dimensions of $\chi=50$, while for estimating the central charge we used matrices up to $\chi=200$.

 \begin{figure}[h]
\includegraphics[width=8cm]{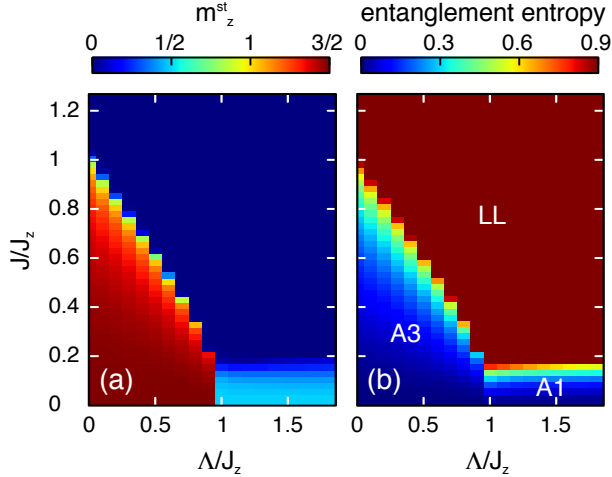}
\caption{(color online) Zero--field phase diagram for the infinite chain as a function of $\Lambda/J_z$ and $J/J_z$, as obtained from DMRG calculation with $\chi=25$. Panel (a) shows the staggered magnetization. Panel (b) shows the half chain entanglement entropy, i.e., the von-Neumann entropy of the reduced density matrix for a bipartition of the chain into two half chains.}
\label{fig:1Dh0PD}
\end{figure}
 
The zero magnetic field phase diagram is shown in Fig.~\ref{fig:1Dh0PD}. We can clearly identify the gapped $A3$ and $A1$ uniaxial phases with finite value of the staggered magnetization $m_z^{\text{st}}$ and small entanglement entropy, and a gapless phase with algebraic correlations (Luttinger liquid). The extension of the $A1$ phase is limited to $J/J_z \alt 0.25$ values, following the estimate based on the mapping to the XXZ model in Sec.~\ref{sec:mappingXXZ}.

\begin{figure}[h]
\includegraphics[width=8cm]{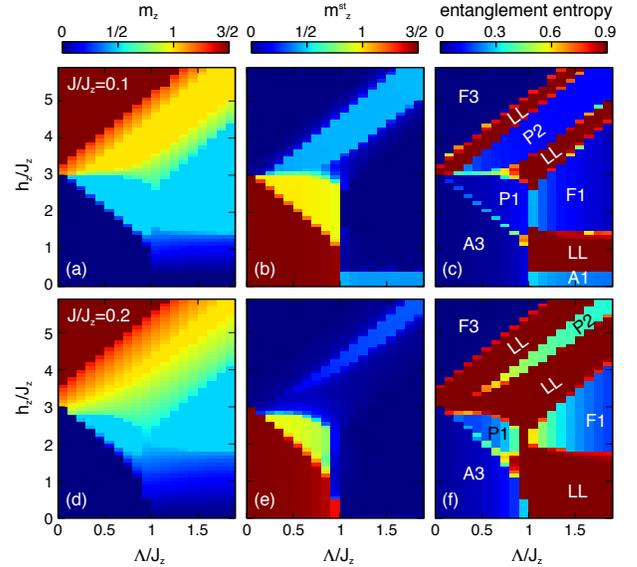}
\caption{(color online) Phase diagram as the function of $\Lambda/J_z$ and $h/J_z$ for (a-c) $J/J_z = 0.1$ and (d-f) $J/J_z = 0.2$. We show the uniform and the staggered magnetization along the $z$ axes, where the plateau phases can be identified. The large increase of the entanglement entropy indicates gapless phases. The phase diagram is obtained from DMRG calculation with $\chi=25$.}
\label{fig:1DPD}
\end{figure}

\begin{figure}[h!]
\begin{center}
\includegraphics[width=8cm]{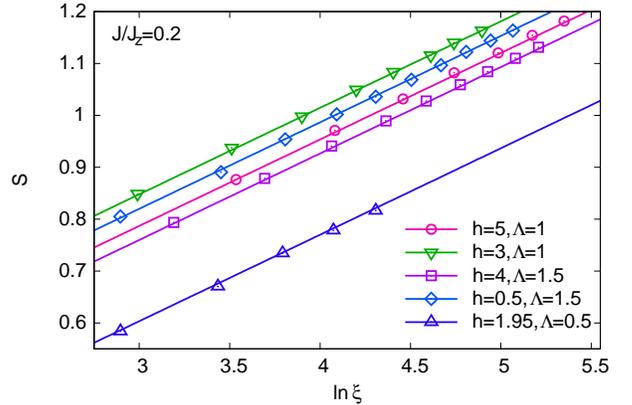}
\caption{(color online) Estimate of the central charge from the entanglement entropy for four different points in the LL phase and one point in the supersolid phase (the $h/J_z=1.95$, $\Lambda/J_z = 0.5$ point). The solid lines are fits based based on the $S=\frac{c}{6} \ln \xi + \text{const.}$ formula, with $c$ set to 1. In all these points the gapless phases are characterized by $c=1$ central charge.}
\label{fig:1Dc}
\end{center}
\end{figure}

 Fig.~\ref{fig:1DPD} shows the phase diagram in the present finite magnetic field. Again, the gapped phases can be identified using the uniform and staggered magnetization ($m_z$ and $m_z^{\text{st}}$), and the small entanglement entropy. The extension of the gapped phases essentially follows the variational phase diagram (see Fig.~\ref{fig:ising2}(b)). However, the supersolid phase is more fragile in the one--dimensional case due to strong quantum fluctuations. Consequently, the gapless phase in the phase diagram is predominantly a simple Luttinger liquid with algebraically decaying correlations and characterized by the integer central charge that measures the number of gapless modes. We calculated the central charge of the gapless phases using the method outlined in Ref.~[\onlinecite{Pollmann2009}] for a few selected points in the phase diagram. Within numerical accuracy we find $c=1$, as shown in Fig.~\ref{fig:1Dc}. This is in accordance with our expectation originating from the mapping to the effective $XXZ$ model, that the gapless phases between the $F3$ and $P2$, $P2$ and $F1$, and $F1$ and $A1$ are all Luttinger liquids. 

\begin{figure}[h!]
\begin{center}
\includegraphics[width=8.5cm]{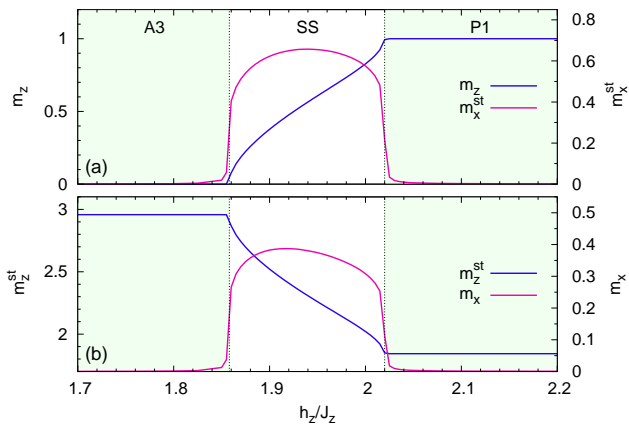}
\caption{(color online) The magnetic field dependence of the order parameters as a function of the magnetic field for $J/J_z = 0.2$ and $\Lambda/J_z = 0.5$. The curves are result of DMRG calculations with $\chi=40$ and with a small field $h_x/J_z = 10^{-4}$. In (a),  the non-vaninshing off--diagonal order parameter $m_x^{\text{st}}$ 
shows the extension of the gapless phases. The finite value of the $m_z^{\text{st}}$ and $m_x$ indicate a robust supersolid phases, as seen in (b).
}
\label{fig:1Dcut}
\end{center}
\end{figure}

\begin{figure}[h!]
\begin{center}
\includegraphics[width=8.5cm]{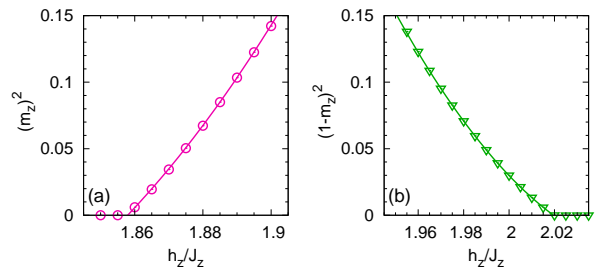}
\caption{(color online) The magnetization has a square root singularity at (a) lower $h_{c,1}/J_z=1.8579$, and (b) upper $h_{c,2}/J_z=2.0195$ critical field. The solid lines show the   
$m_z^2 \approx 2.68 (h-h_{c,1})/J_z+16.9 (h-h_{c,1})^2/J_z^2$ and $(1-m_z)^2 \approx 1.27 (h_{c,2}-h)/J_z+13.2 (h_{c,2}-h)^2/J_z^2$ fits to the magnetization curves.}
\label{fig:1DsingSS}
\end{center}
\end{figure}

We have searched for a supersolid phase in the vicinity of the gapped phases that break the translational invariance. We  made a scan by varying the field for a fixed value of $\Lambda/J_z$ and $J/J_z$; the results are plotted in Fig.~\ref{fig:1Dcut}. For the simulations,  we added a tiny magnetic field of order $10^{-4}$ along the $x$-axis to break the U(1) symmetry around the-$z$ axis. A finite value of the diagonal (staggered magnetization $m_z^{\text{st}}$) and off--diagonal (magnetization along the $x$ axis, $m_x$) order parameter indicates the presence of the supersolid.
It appears that the supersolid  is stable in a small region only, between the $A3$ and $P1$ gapped phase, with a continuous phase transitions. Both the magnetization and the staggered magnetization in the supersolid show a square root like behavior at the lower and upper critical fields, like the magnetization in XXZ model does, see for example in Ref.~[\onlinecite{PhysRevLett.101.137207}]. This is due to the density of the states of the spinons, and is already observed in the $XY$ model, when it is
mapped to free fermions. Recall that the density of states of free fermions has a van Hove singularity at the band edges, and this shows up as a square root singularity in the magnetization curve of XY (and XXZ) model.  
In Fig.~\ref{fig:1DsingSS} we straighten out this singularity. This singularity is also inherited for the staggered magnetization at the critical fields. The central charge in the supersolid is also $c=1$ (the lowest line in Fig.~\ref{fig:1Dc}, here we set the $h_x =0$, as otherwise a finite $h_x$ induces a gap in the spectrum). 

From variational calculations, we expect a continuos phase transitions into the supersolid also at the upper edge of the $P_1$ phase. Numerically, however, we find a first order transition into the LL phase.

\section{Exact Diagonalization studies}\label{sec:exactdiag}

To get further insight into the problem in higher dimensions, we have numerically diagonalized  small (8- and 10-site) clusters of  spin $S=3/2$ arranged on the square lattice with periodic boundary condition and  searched for signature of different phases in the energy spectrum. The two--sublattice states break translation symmetry, so we expect two degenerate ground states with momentum $\mathbf{k} = (0,0)$ and $(\pi,\pi)$ in the thermodynamic limit. In the gapped phases, these two levels are well separated from the other states, while in the supersolid, where both translational symmetry and U(1) symmetry breaking occurs, we expect two copies of the Anderson towers in the spectrum that is the signature of the $U(1)$ symmetry breaking.\cite{Anderson1952, Bernu1992} Unfortunately, the large spin makes the finite size scaling difficult, and without a finite size scaling we cannot be sure about the exact nature of the ground state. Nevertheless, even our small cluster gives important support for the variational phase diagram. 

\begin{figure}[bth]
\includegraphics[width=8cm]{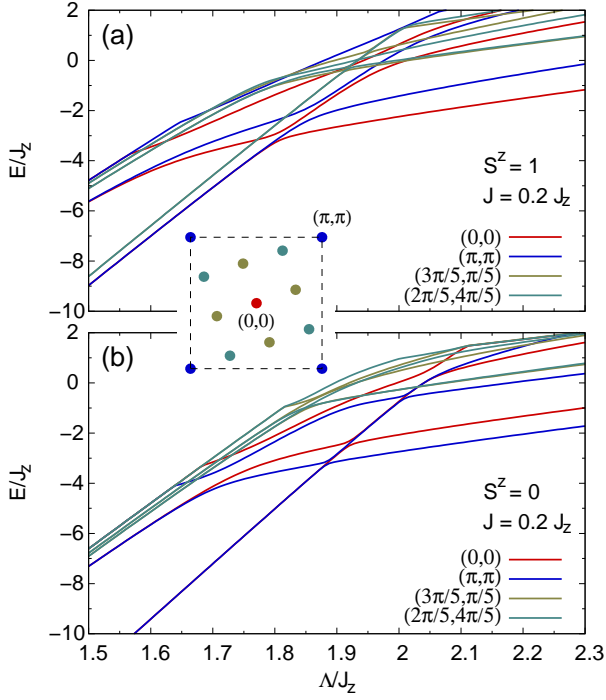}
\caption{(color online) The first few lowest lying energy levels of a 10 site cluster for (a) $S^z=0$ and (b) $S^z=1$ as a function of $\Lambda/J_z$. We set $J = 0.2 J_z$. The inset shows the available $\mathbf{k}$--points in the Brillouin zone.}
\label{fig:ED_spectrum}
\end{figure}

 In Fig.~\ref{fig:ED_spectrum} we show the energy spectrum for the $C_4$ symmetric 10 site cluster and $J = 0.2 J_z$ 
 around $\Lambda = 2 J_z$, where we expect the first order transition from the phase $A3$ into the supersolid to happen. In zero field the ground state has $S^z=0$, and in Fig.~\ref{fig:ED_spectrum}(b) we see that the energy level curvatures of lowest lying states in the $\mathbf{k} = (0,0)$ and $(\pi,\pi)$ sector are essentially indistinguishable for $\Lambda \alt 1.88 J_z$ and well separated from the higher levels. This indicates the presence of a gapped, two--sublattice state that we can associate with the $A3$ phase. The sharp level anti-crossing  at  $\Lambda \approx 1.88 J_z$ indicates a first order transition. In the $S^z=1$ sector we observe the spin excitations with a narrow bandwidth and a $\mathbf{k} \leftrightarrow (\pi,\pi)- \mathbf{k}$ symmetry, following Eq.~(\ref{eq:wA3P1}) as calculated from the perturbation theory in Sec.~\ref{sec:ising_2sublattice_pert}. For $\Lambda \agt 1.88 J_z$, the $\mathbf{k} = (0,0)$ and $(\pi,\pi)$ levels are also  close, and the these two levels are equally close and reversed in order for $S^z=1$ in Fig.~\ref{fig:ED_spectrum}(a), an indication for the U(1) symmetry breaking, possibly with translational symmetry breaking (the supersolid phase).

\begin{figure}[bth]
\includegraphics[width=7cm]{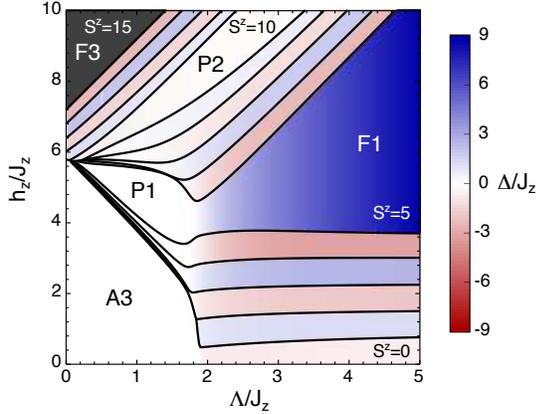}
\caption{(color online) The gap $\Delta = E_{(\pi,\pi)}-E_{(0,0)}$ as a function of $\Lambda/J_z$ and $h/J_z$ for the 10--site cluster. The solid curves separate the different $S^z$ sectors.}
\label{fig:ED_gapPD}
\end{figure}

\begin{figure}[bth]
\begin{center}
\includegraphics[width=8cm]{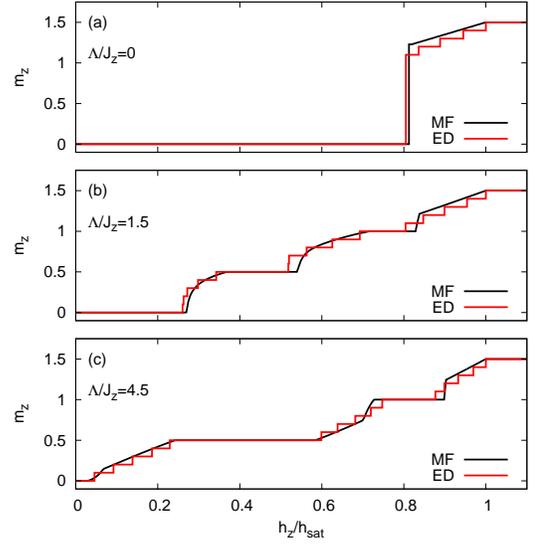}
\caption{(color online) Magnetization as a function of magnetic field, as obtained from variational calculation and exact diagonalisation. Here $J=0.2 J_z$, and $\Lambda/J_z=$0, 1.5, and 4.5.  $h_{\text{sat}}$ is the saturation field [Eq.~(\ref{eq:hsat})]. }
\label{fig:magnetizationEDvari}
\end{center}
\end{figure}

 In Fig.~\ref{fig:ED_gapPD} the energy gap between the $\mathbf{k} = (0,0)$ and $(\pi,\pi)$ ground states in the different $S^z$ sectors is shown as a function of 
 $\Lambda/J_z$ and magnetic field, as this may serve as an indicator of the translational symmetry breaking. We can identify the gapped phases (except for $A1$) and their extension is even quantitatively in good agreement with the variational phase diagram, shown in Fig.~\ref{fig:ising2}(b). The consistency between the variational and exact diagonalization result is also supported in Fig.~\ref{fig:magnetizationEDvari}, where we compare the magnetization calculated by these two methods.

\section{Conclusions}
\label{sec:concusions}

In this paper we studied the effect of exchange and easy--plane anisotropies  on the formation of magnetization plateaus and supersolids in spin--3/2 system on (unfrustrated) bipartite lattices, with the aim to extend the results of earlier studies on the stability of supersolids in anisotropic  spin--1  model on the square lattice\cite{Sengupta2007} and spin--1/2 bilayer systems \cite{Ng2006,Laflorencie2007,Picon2008} to larger values of spins.

 In the Ising limit ($J=0$) we find both uniform and translational symmetry breaking magnetic phases with gapped excitation spectrum with zero of finite magnetization (magnetization plateaus). We discussed the macroscopic degeneracy of the ground state at the phase boundaries and showed that when the off--diagonal exchange interaction $J$ becomes finite this degeneracy is lifted and new gapless phases emerge. All the plateaus continuously evolve from the Ising limit, and the degeneracy of the boundaries in the Ising--limit gives a hint on the order of the phase transition and on the nature of the gapless state. Not surprisingly, our variational calculation shows that the supersolid phases are concentrated around the plateaus that break the translational symmetry. In particular, the tendency toward supersolidity is greatly enhanced when the degeneracy of the boundary is $2 \times 2^{N/2}$ (due to the choice of two states on the sites of one of the sublattices, while the sites of the other sublattice are occupied with a third type of states), as in this case the diagonal translational order is preformed, and the off--diagonal order is easily established on the sublattice occupied with the two states. In addition, for large anisotropies we have confirmed the stability of the plateau states using perturbation theory, and found a good agreement between the two approaches regarding the extension of the gapped phases.

In zero field we plotted the variational phase diagram as function of the on--site and exchange anisotropies. Aside from the axial antiferromagnetic phases and planar superfluid phase, we find a biconical superfluid which simultaneously exhibits the diagonal and off--diagonal staggered characteristics of the former phases.

In the $J=J_z$ Heisenberg limit, when the exchange interaction is SU(2) invariant (but we keep the on--site anisotropy $\Lambda$ that breaks the SU(2) symmetry), the plateau and supersolid phases disappear and only the uniform phases and the superfluid phase between them are present.

The variational phase diagrams for zero and finite magnetic field  were compared to DMRG calculations carried out in one dimension. We have found convincing evidence for a supersolid state that is realized in a region between two gapped phases that break the translational symmetry. For a two-dimenal square lattice, we performed exact diagonalisation in two dimension on small clusters for $J/J_z=0.2$ and identified the characteristics of  various phases from the energy spectrum. The extension of the gapped phases based on these calculations proved to be in good qualitative and quantitative agreement with the variational findings. 

Our study was initially inspired by the Ba$_2$CoGe$_2$O$_7$ layered material, where Ref.~\onlinecite{Miyahara2011} estimates $\Lambda/J_z\approx 8$ and $J\approx J_z$. While these anisotropies are not strong enough to stabilize a magnetization plateau, an anomaly occurs around $m/m_{\text{sat}} = 1/3$  in the magnetization curve -- this is also observed experimentally: the magnetization curve changes it's slope at $h\approx 9$T.\footnote{H. Murakawa, private communication}

\begin{acknowledgments}
We are pleased to thank stimulating discussions with S.~Bord\'acs, I.~K\'ezsm\'arki, L.~Seabra and P.~Sindzingre. We are also grateful for H. Murakawa for sharing his magnetization measurements with us. This work was supported by Hungarian OTKA Grant Nos. K73455 and NN76727, and the guest program of MPI Physik Komplexer Systeme in Dresden.
\end{acknowledgments}

\appendix

\section{Perturbation expansion}
\label{sec:perturbation}
%

Here we are presenting the results of the Rayleigh-Scr\"odinger perturbation theory applied to states and excitations in the $J \rightarrow 0$ limit.

\subsection{Second order corrections in J to the ground-state energy}

The second order correction to the energy/(per site) of the different phases are as follows:
\begin{eqnarray}
\varepsilon_{A1}^{(2)} & = & - \frac{8 J^2}{3 J_z} - \frac{9 J^2}{2(4 \Lambda - 5 J_z)}  \label{eq:e2A1}\\
\varepsilon_{A3}^{(2)} & = & - \frac{9 J^2}{2(11 J_z - 4 \Lambda)} \label{eq:e2A3}\\
\varepsilon_{F1}^{(2)} & = & - \frac{12 J^2}{2 \Lambda - J_z} \label{eq:e2F1}\\
\varepsilon_{P1}^{(2)} & = & - \frac{6 J^2}{7 J_z-2 \Lambda} \label{eq:e2P1}\\
\varepsilon_{P2}^{(2)} & = & - \frac{3 J^2}{2 J_z} \label{eq:e2P2}\\
\varepsilon_{F3}^{(2)} & = & 0 \label{eq:e2F3}
\end{eqnarray}

\subsection{First order degenerate perturbation theory for excitation spectrum of the uniform F1 and F2 phases}
\begin{eqnarray} 
\omega_{F1 \rightarrow P2} &=& -h + 2 J_z + 2 \Lambda + 6 J \gamma_{\mathbf{k}} \label{eq:wF1P2}
\\
\omega_{F1 \rightarrow A1} &=& h - 2 J_z + 8 J \gamma_{\mathbf{k}} \label{eq:wF1A1}
\\
\omega_{F3 \rightarrow P2} &=& h - 6 J_z - 2 \Lambda + 6 J \gamma_{\mathbf{k}} \label{eq:wF3P2}
\end{eqnarray}
where $\gamma_{\mathbf{k}}$ is defined in Eq.~\ref{eq:gammak}.
 
\subsection{Second order degenerate perturbation theory for excitation spectrum of the staggered phases}

\begin{widetext}
\begin{eqnarray} 
\omega_{P1 \rightarrow A3} &=& h + 2 \Lambda - 6 J_z
- \frac{36 J^2}{8 J_z - 2 \Lambda}
- \frac{9 J^2}{4( 8 J_z - 4 \Lambda) } 16\gamma_{\mathbf{k}}^2 +\frac{48 J^2}{7 J_z-2 \Lambda }
\label{eq:wP1A3}
\\
\omega_{P1 \rightarrow P2} &=& -h + 6 J_z -\frac{3 J^2}{2 J_z}+\frac{48 J^2}{7 J_z-2 \Lambda }-\frac{3 J^2}{6 J_z-2 \Lambda } (16 \gamma^2_{\mathbf{k}}+8)
\label{eq:wP1P2}
\\
\omega_{A1 \rightarrow F1} &=& -h + 2 J_z-\frac{27 J^2}{4\Lambda-4J_z}-\frac{12 J^2}{2\Lambda-2 J_z} 
+\frac{64 J^2}{3 J_z} + \frac{36 J^2}{4 \Lambda - 5 J_z} 
-\frac{2 J^2}{J_z} (16 \gamma^2_{\mathbf{k}}+8)
- \frac{3 J^2}{ 2 \Lambda - 4 J_z }16\gamma_{\mathbf{k}}^2
\label{eq:wA1F1}
\\
\omega_{P2 \rightarrow F3} &=& -h + 6 J_z+2 \Lambda -\frac{9 J^2}{8 J_z}(16 \gamma^2_{\mathbf{k}}+8) + 12 \frac{J^2}{J_z}
\label{eq:wP2F3}
\\
\omega_{P2 \rightarrow F1} &=& h - 2 J_z - 2 \Lambda - \frac{12J^2}{2 J_z + 2 \Lambda} - \frac{9 J^2}{ 8 J_z }16 \gamma^2_{\mathbf{k}} 
- \frac{3 J^2}{2\Lambda-4 J_z } 16 \gamma^2_{\mathbf{k}}
+ \frac{3 J^2}{ J_z }
\label{eq:wP2F1}
\\
\omega_{P2 \rightarrow P1} &=& h - 6 J_z + \frac{21 J^2}{4 J_z} - \frac{3 J^2}{4 J_z - 2\Lambda } 16 \gamma^2_{\mathbf{k}}
\label{eq:wP2P1}
\\
\omega_{A3 \rightarrow P1} &=& -h + 6 J_z-2 \Lambda
- \frac{12 J^2}{10 J_z - 2 \Lambda } + \frac{36 J^2}{11 J_z - 4 \Lambda } -\frac{9 J^2}{8 (5 J_z - 2 \Lambda )}(16 \gamma^2_{\mathbf{k}}+8)
\label{eq:wA3P1}
\end{eqnarray}

In case that we include the excitations on both sublattices, the $S_i^-$ excitations from the A1 in the $\mathbf{k}$ space are eigenvalues of the  
\begin{equation}
\mathcal{H}_{A1} = 
\left(
\begin{array}{cc}
2 J_z-h-\frac{2 J^2}{J_z} (16 \gamma^2_{\mathbf{k}}+8)-\frac{27 J^2}{4\Lambda-4J_z}-\frac{12 J^2}{2\Lambda-2 J_z}
&  
4 \sqrt{3} J \gamma_{\mathbf{k}}   
\\
4 \sqrt{3} J \gamma_{\mathbf{k}}   
&     
2 \Lambda - 2 J_z - h - \frac{12J^2}{J_z} -\frac{9 J^2}{4 (4\Lambda - 6 J_z)}(16 \gamma^2_{\mathbf{k}}+8) 
\end{array}
\right) - 8 \varepsilon_{A1}^{(2)}
\label{eq:H2x2A1}
\end{equation}
matrix. If we expand for $J$ up to second order, this will give Eq.~\ref{eq:wA1F1}, the corrections to the dispersion directly to the F1 phase.

Similarly, for the P1 phase
\begin{equation}
\mathcal{H}_{P1} = 
\left(
\begin{array}{cc}
2 \Lambda + h - 6 J_z - \frac{36 J^2}{8 J_z - 2 \Lambda}
&  
6 J  \gamma_{\mathbf{k}}   
\\
6 J \gamma_{\mathbf{k}}   
&     
2 J_z + h - 2 \Lambda - \frac{8J^2}{3 J_z} - \frac{3 J^2}{ 6 J_z - 2 \Lambda}(16 \gamma^2_{\mathbf{k}}+8) 
\end{array}
\right) - 8 \varepsilon_{P1}^{(2)} ,
\label{eq:H2x2P1}
\end{equation}
and for the P2 phase:
\begin{equation}
\mathcal{H}_{P2} = 
\left(
\begin{array}{cc}
- 6 J_z + h - \frac{27 J^2}{4 J_z}
&  
4 \sqrt{3} J \gamma_{\mathbf{k}}   
\\
4 \sqrt{3} J \gamma_{\mathbf{k}}   
&     
- 2 J_z  + h - 2 \Lambda - \frac{12J^2}{2 J_z + 2 \Lambda} - \frac{9 J^2}{ 8 J_z }(16 \gamma^2_{\mathbf{k}}+8) 
\end{array}
\right) - 8 \varepsilon_{P2}^{(2)} .
\label{eq:H2x2P2}
\end{equation}


\end{widetext}



\bibliographystyle{unsrt}
\bibliographystyle{apsrev4-1}
\bibliography{Ba2CoGe2O7_ToSubmit}
\input{Ba2CoGe2O7_ToSubmit.bbl}
\end{document}